\documentclass[aps,prl,twocolumn,groupedaddress]{revtex4-1}
\usepackage{graphicx}
\usepackage{bm}
\usepackage{amsmath,amssymb}
\usepackage{epstopdf} 

\newcommand{\BS}{Bi$_2$Se$_3$}

\newcommand{\Tc}{$T_{\rm{C}}$}

\renewcommand{\vec}[1]{\mbox{\boldmath$1$}}

\newcounter{lastnote}

\def\bc{\begin{center}}
\def\ec{\end{center}}
\def\be{\begin{equation}}
\def\ee{\end{equation}}
\renewcommand{\vec}[1]{\mbox{\boldmath$1$}}

\begin{document}

\title{Growth and Characterization of Hybrid Insulating Ferromagnet-Topological Insulator Heterostructure Devices}

\author{A. Kandala}
\author{A. Richardella}
\author{D. W. Rench}
\author{D. M. Zhang}
\altaffiliation{Current address: Center for Nanoscale Science and Technology, National Institute of Standards and Technology, Gaithersburg, MD 20899, USA.}
\author{T. C. Flanagan}
\author{N. Samarth}
\email{nsamarth@psu.edu}
\affiliation{Department of Physics and Materials Research Institute, The Pennsylvania State University, University Park, PA 16802, USA}

\date{\today}

\begin{abstract}
We report the integration of the insulating ferromagnet GdN with epitaxial films of the topological insulator \BS~ and present detailed structural, magnetic and transport characterization of the heterostructures. Fabrication of multi-channel Hall bars with bare and GdN-capped sections enable direct comparison of magnetotransport properties. We show that the presence of the magnetic overlayer results in suppression of weak anti-localization at the top surface.
\end{abstract}


\maketitle

Combining three dimensional (3D) topological insulators (TIs) with magnetism is of great current interest because of the unique effects predicted when the time reversal protected topological surface states are modified by symmetry breaking magnetic perturbations \cite{Hasan2010,Qi2011}. Experiments aimed at observing such phenomena have primarily focused on magnetically doped TIs  \cite{Liu2012,Checkelsky2012,Xu2012a,Zhang2012,Chang2013}. Complementary to the direct magnetic doping of TIs, several theoretical schemes have proposed explorations of a different sample geometry wherein a patterned ferromagnet (FM) is interfaced with a TI \cite{Mondal2010,Garate2010,Kong2011,Tserkovnyak2012}. When the magnetic easy axis of the FM overlayer is out-of-plane, a gap opens in the surface states of the vicinal TI and this massive Dirac Hamiltonian should lead to chiral 1D edge states along FM domain walls where the mass changes sign. If the easy axis of the FM lies in-plane, no gap is expected to first order but the moment can be levered out-of-plane by an external field, thereby opening or closing a gap. Calculations\cite{Oroszlany2012} also show that hexagonal warping effects can lead to a gap even with an in-plane FM. Unlike magnetically doped films \cite{Liu2012}, such a geometry leaves the bulk band structure unaffected. However, electrical transport experiments require an insulating ferromagnet to ensure current paths that flow solely through the TI. Recent advances in this context have exploited the synthesis of EuS/\BS~heterostructures \cite{Wei2013,Yang2013}. In this {\it Letter}, we demonstrate an alternative scheme towards such ``magnetic gating'' proposals by creating hybrid electrical transport devices wherein we interface the insulating FM GdN with a TI (\BS). 

GdN is an insulating FM that has elicited interest for low temperature spintronic devices, particularly since it can be deposited by reactive sputtering at ambient temperature \cite{Xiao1996,Senapati2011b}. This is of prime importance in studies of magnetic exchange coupling effects, as it minimizes thermal diffusion of magnetic species. Changing the nitrogen composition allows access to metallic \cite{Plank2011}, semiconducting \cite{Granville2006} and insulating \cite{Senapati2011b} regimes, opening possibilities for its use as a spin-injector/detector, a spin-filter or a ferromagnetic gate in novel spintronic devices. The semiconducting and insulating states display a maximum ferromagnetic Curie temperature \Tc $\sim 65-70$ K \cite{Senapati2011b,Plank2011,Granville2006}, while more conducting N-deficient forms can have \Tc~well above 100 K \cite{Senapati2011b,Plank2011}.
 
\BS~thin films were grown by molecular beam epitaxy (MBE) on InP (111)A substrates under typical growth conditions described elsewhere \cite{Richardella2010}. Following a brief exposure to ambient atmosphere, the films were transferred to a Kurt Lesker CMS-18 system. After an {\it in-situ} Ar$^+$ surface clean, GdN was deposited by reactive rf sputtering of Gd in an Ar:N$_2$ environment at ambient temperature. The GdN films were deposited at a rate of $\sim 0.1-0.2$ $\AA$/sec, with a sputtering power of 4.93 W/cm$^2$ in a 15\%~N$_2$:Ar gas environment at 5 mTorr pressure with a source to substrate distance of 20 cm. The base pressure in the chamber prior to sputtering was $\sim$ 4x10$^{-7}$ Torr. The lack of ultrahigh vacuum is probably responsible for the lower \Tc ($\sim 13$K) of our films. Since GdN oxidizes instantly on exposure to atmosphere, the films were capped {\it in-situ} with $\sim$ 60 nm of Au deposited by dc sputtering.

\begin{figure}[t]
\includegraphics[width=3.5in]{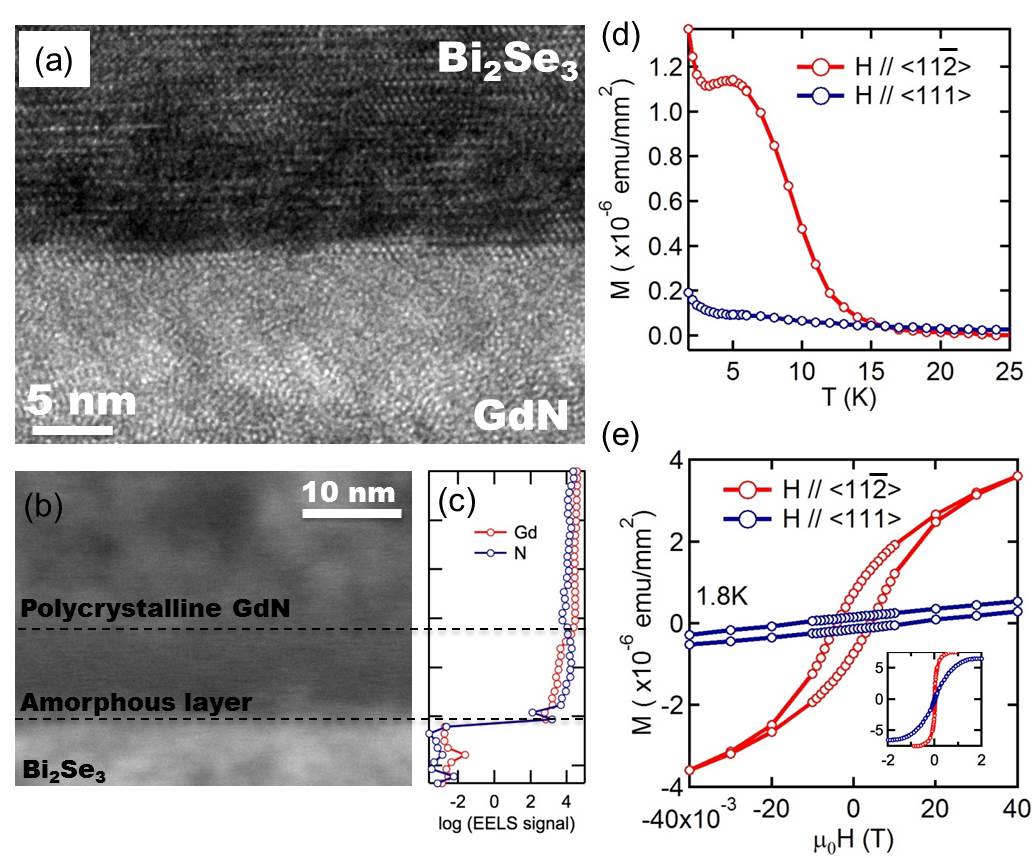} 
\caption{(Color online) \text{(a)} HRTEM phase contrast image of the heterostructure. \text{(b)} STEM image of the heterostructure. Dotted lines differentiate the \BS~ film, an interfacial amorphous GdN layer, and the polycrystalline GdN layer. \text{(c)} Corresponding line scan of the Gd (M4,5 edge) and N (K edge) EELS signal reveals a sharp drop at the \BS~ interface. No Gd signal is detected in the bulk of the \BS. \text{(d)} $M$ vs $T$ measurements of the \BS/GdN heterostructure reveal \Tc $\sim$ 13 K. \text{(e)} $M$ vs $\mu_0 H$ measurements at $T = 1.8$ K along the $\langle 11\bar{2} \rangle$ and $\langle111 \rangle$ directions of the substrate, showing that the easy axis is in-plane. Inset: $M$ vs $\mu_0 H$ curves over a 2 T field range.}
\label{TEM}
\end{figure}

Structural and magnetic characterization of the heterostructure is presented in Fig. 1. High-resolution transmission electron microscopy (HRTEM) and electron energy loss spectroscopy (EELS) were used to investigate the nature of the interface. The phase contrast image in Fig. 1(a) reveals a sharp interface between the epitaxial \BS~ and the GdN. However, bright field images (see Fig. S1 in supplementary materials) reveal that the first 5-10 nm of the GdN is amorphous, followed by polycrystalline GdN thereafter. Scanning TEM (STEM) images and corresponding Gd EELS intensities are shown in Figs. 1(b) and (c). The line scan of the energy loss near-edge structure for the Gd M4,5 edge, obtained using a probe size of 0.5 nm and step size of 0.72 nm, shows no Gd in the bulk of the \BS. We emphasize that the lack of Gd diffusion into the bulk of the \BS~clearly differentiates this scheme from previous studies of bulk magnetically doped TIs. More detailed structural characterization may be found in the supplementary section. Magnetization measurements in a superconducting quantum interference device (SQUID) magnetometer confirmed the ferromagnetism in the heterostructures. Measurements were taken along both in-plane and perpendicular-to-plane directions in a 5 mT measuring field after a 0.9 T field cool. The temperature dependence of the magnetization $M$ shows  \Tc~ $\sim$13 K (Fig. 1(d)). The hysteresis loops in the $M$ v/s $\mu_0H$ plots (Fig. 1(e)) show that the easy axis is in-plane, possibly due to shape anisotropy of the GdN thin film. 

We measured magneto-transport in Hall bar devices of an 8 nm thick \BS~thin film, patterned into multiple channels of dimension 20 $\mu$m $\times$ 60 $\mu$m by standard photolithography and etching. The ferromagnetic GdN/Au gate electrode is defined over one of the channels by another step of photolithography, sputtering and a lift-off process. The sputtering conditions were identical to those employed for the characterization in Fig. 1. Figure 2(a) is an image of a typical device showing channels of bare \BS~and ones covered with GdN. This enables a comparison of the magneto-transport properties of the \BS~film in both regions to see the effects of the overlying magnetic layer. DC transport measurements were carried out in a Quantum Design PPMS system with an 8 T superconducting magnet and a base temperature of 1.8 K and an Oxford Heliox He$^{3}$ system with a base temperature of 400 mK. A typical non-linear  2-point {\it I-V} curve of the Au-GdN-\BS~structure, is shown in Fig. 2(b), indicative of a barrier to transport. To confirm that the barrier indeed lies in the GdN and that the GdN is itself insulating, we additionally patterned Hall bars from a single \BS~film but with varying thickness of the overlying GdN layer. The room temperature resistance of the capped channel in these devices showed no marked change for different thicknesses of the GdN. This is shown in the 4-pt {\it V-I} curves in Fig. 2(c) for GdN thicknesses ranging from 15-30nm. These measurements rule out the possibility of alternate current paths through the GdN layer, and demonstrate that charge transport is solely restricted to underlying \BS.

\begin{figure}[t]
\includegraphics[width=3.5in]{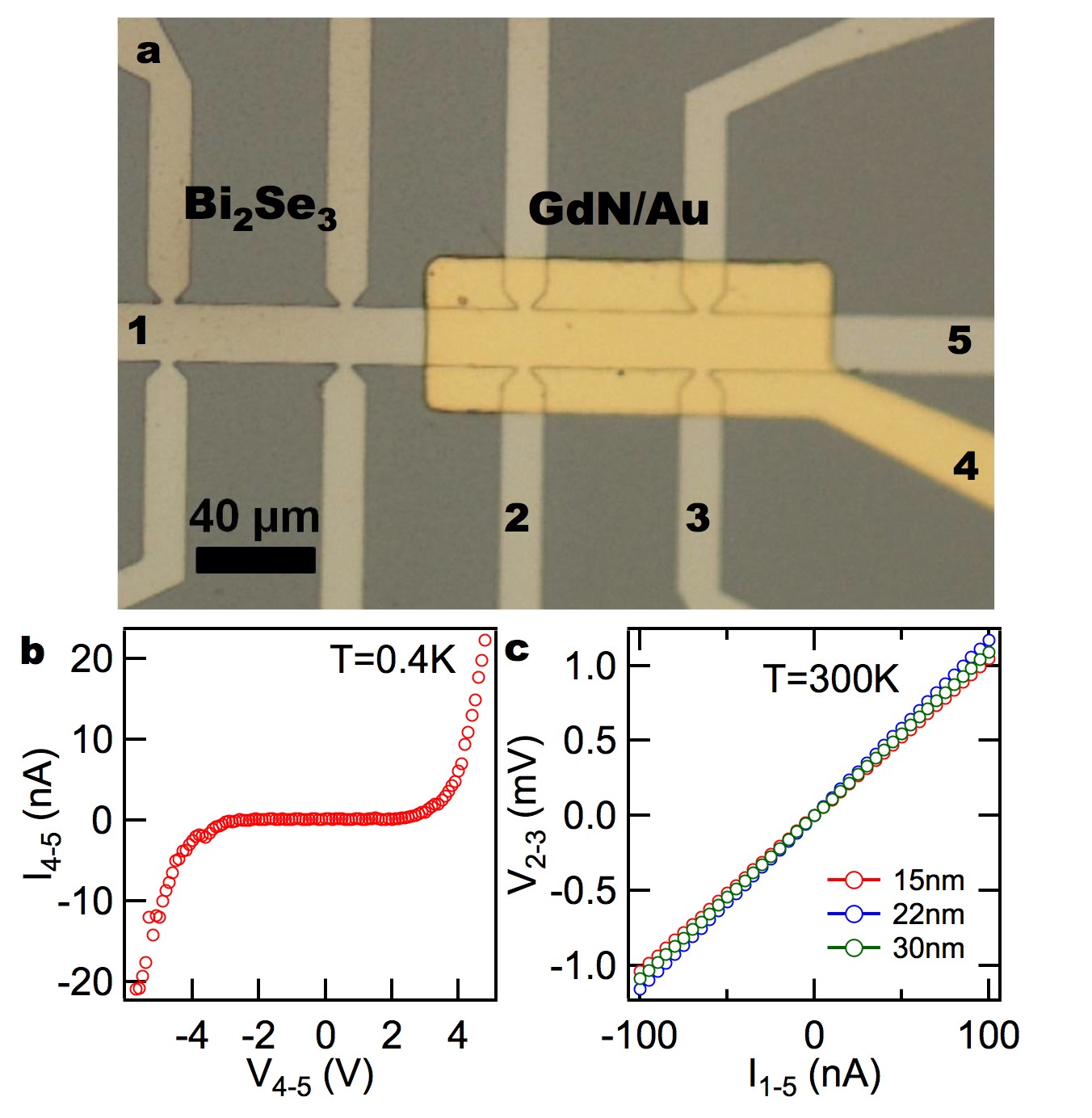} 
\caption{(Color online) \text{(a)} Optical microscope image of the transport device. \text{(b)} Non-linear 2-point {\it I-V} curve of the Au-GdN-\BS~ structure at $T = 400$ mK shows a strong non-linearity indicative of a barrier to transport.\text{(c)} 4-point {\it V-I} curves of GdN capped channels patterned from the same thin film for different GdN thicknesses: 15 nm (red), 22 nm (blue) and 30 nm(green). The channel resistance shows no marked change with increasing GdN thickness, confirming the insulating nature of the GdN.}

\end{figure}

A comparison of the transport properties of the two channels in the device directly reveals the influence of the overlying magnetic layer on transport in \BS. The resistances of both channels show metallic temperature dependences (Fig 3(a)), except for an upturn at low temperatures where the two dimensional (2D) sheet conductance ($\sigma$) has a logarithmic temperature dependence $\sigma\sim\ln (T)$ (inset to Fig. 3(a)), consistent with electron-electron (e-e) interaction in 2D~ \cite{Wang2011}. The smaller residual resistivity ratio of the capped channel seen in Fig. 3(a) is already indicative of its lower mobility, and further confirmed by Hall measurements. The bare channel has a resistivity $\rho = 1.17$ m$\Omega$.cm, mobility $\mu =$ 246 cm$^{2}/$V.s and carrier density $n = 2.17 \times 10^{19}$ cm$^{-3}$; the capped channel has $\rho = 4.12$ m$\Omega$.cm, $\mu =$ 60.9 cm$^{2}/$V.s and carrier density $n = 2.49 \times 10^{19}$ cm$^{-3}$. One potential cause for this suppression in mobility is damage from sputtering by reflections of energetic Ar gas neutrals, as observed in graphene \cite{Chen_Arxiv}.

The carriers are n-type, and their density corresponds to a Fermi energy in the bulk conduction band. Figure 3(b) compares the low-temperature magneto-conductance MC in the two channels over a temperature range $2 \leq T \leq 10$K. Both channels show a negative MC that is characteristic of weak anti-localization (WAL) in \BS~due to strong spin-orbit coupling in the bulk and surface states. We note that low-frequency (19 Hz) lock-in measurements in the capped channel display a positive MC which only depends on the perpendicular component of magnetic field (data not shown). This positive MC might be readily interpreted as evidence for weak localization arising from opening of a surface state gap \cite{Lu2011a}. However, we find that the positive MC is frequency-dependent and is absent in the DC measurements reported in this {\it Letter}. Thus, we cannot attribute it to the opening of a gap and the phenomenon is not currently understood. Our observations contrast with measurements in EuS/\BS~heterostructures where positive MC signatures of gap-induced weak localization are seen in both DC and AC transport \cite{Yang2013}.) 

We extract the phase coherence lengths $l_{\phi}$ in the two channels by fitting the MC using the Hikami-Larkin-Nagaoka equation \cite{Hikami1980} for WAL quantum corrections, ignoring for simplicity the effects of e-e interactions \cite{Wang2011}:
\begin{equation}
\Delta\sigma(B)= \frac{\alpha e^2}{\pi h}\left[\Psi\left(\frac{l^2_B}{l^2_{\phi}}+\frac{1}{2}\right)-\ln \left(\frac{l^2_B}{l^2_{\phi}}\right)\right].
\end{equation}
In the above equation $\alpha$ and $l_{\phi}$ are used as fitting parameters and $l_B$ is the magnetic length. In the relevant case of strong spin-orbit coupling and weak magnetic scattering, we expect that $\alpha \sim -0.5$ for a single coherent channel, and the system lies in the symplectic class. In the limit of strong magnetic scattering, we expect $\alpha \sim 0$, and the system lies in the unitary class. The temperature dependence of $l_{\phi}$ and $\alpha$ are compared for the two channels in Figs. 3(c) and 3(d), respectively. Both capped and bare channels show a power law temperature dependence $l_{\phi} \propto T^{-1/2}$, suggesting that the dominant dephasing mechanism is e-e interactions in 2D. However, the capped channel has a suppressed $l_{\phi}$ in comparison to the bare channel. This might be trivially attributed to magnetic scattering. However, note that the value of $\alpha$ for the capped channel is inconsistent with a transition towards a unitary class due to magnetic scattering, and instead remains in the symplectic class (taking values close to $-0.5$). This raises the possibility that the WAL with the shorter $l_{\phi}$ in the capped channel is associated with a single coherent channel that couples the bulk and the bottom surface state. In such a scenario, contributions to WAL from the top surface state may be expected to be completely suppressed due to the overlying magnetic layer.

\begin{figure}[t]
\includegraphics[width=3.5in]{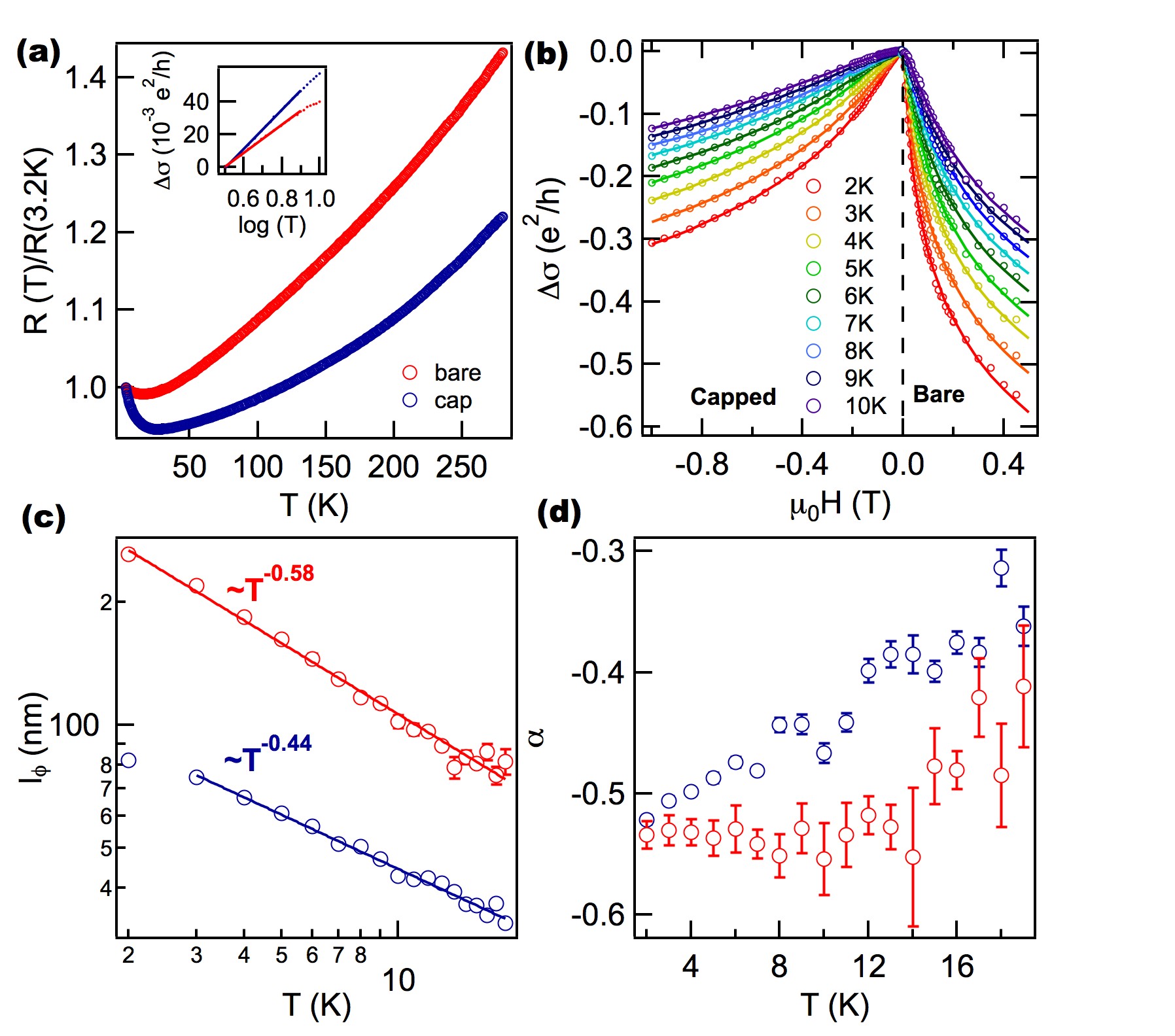} 
\caption{(Color online) \text{(a)} Residual resistivity ratio (RRR) $R(T)/R(3.2K)$ for the GdN capped (blue circles) and bare (red circles) channels. Both show a metallic temperature dependence, and the weaker RRR for the capped channel is indicative of its lower mobility. Inset shows that $\sigma \propto \ln (T)$ at low temperature for both channels due to e-e interactions. \text{(b)} MC for the capped channel (left half of panel) and the bare channel (right half of panel) at temperatures ranging from 2 K to 10 K. Circles are data points and the bold lines are quantum corrections fits to WAL. \text{(c)} Temperature dependence of the phase breaking length $l_{\phi}$ for the capped channel (blue) and bare channel (red) from the fits in Fig. 3(b). Bold lines represent power law fits. \text{(d)} Temperature dependence of the prefactor $\alpha$ for the bare channel (red) and the capped channel (blue).  Both channels take values close to $- 0.5$, and hence lie in the symplectic class. Bars on the data points of (c) and (d) represent 95\% confidence intervals of the fits.}
\end{figure}

In summary, we demonstrated the growth of a novel class of TI heterostructures by interfacing \BS~with the insulating ferromagnet GdN. The ability to deposit GdN at ambient temperatures minimizes Gd diffusion into the \BS, as confirmed by high resolution EELS measurements. Also, electrical characterization of the heterostructure confirms the insulating nature of our GdN films, and thus ensures lateral transport solely through the \BS. Finally, our hybrid devices allow a direct comparison of magneto-transport properties of \BS~channels with and without magnetic interactions. Devices with lower bulk carrier densities TI thin films and further improvements in the interface may enable direct access to the effects of broken time-reversal symmetry. The development of these heterostructure devices opens up a broad range of opportunities for studying the influence of magnetism on electrical transport in the topological surface state and the development of novel spintronic devices.

This work was supported by DARPA (N66001-11-1-4110), ONR (N00014-12-1-0117) and the Pennsylvania State University Materials Research Institute Nanofabrication Lab (ECS-0335765). We are grateful to Ke Wang, Josh Maier and Trevor Clark for assistance with TEM.


\begin{thebibliography}{22}%
\makeatletter
\providecommand \@ifxundefined [1]{%
 \@ifx{#1\undefined}
}%
\providecommand \@ifnum [1]{%
 \ifnum #1\expandafter \@firstoftwo
 \else \expandafter \@secondoftwo
 \fi
}%
\providecommand \@ifx [1]{%
 \ifx #1\expandafter \@firstoftwo
 \else \expandafter \@secondoftwo
 \fi
}%
\providecommand \natexlab [1]{#1}%
\providecommand \enquote  [1]{``#1''}%
\providecommand \bibnamefont  [1]{#1}%
\providecommand \bibfnamefont [1]{#1}%
\providecommand \citenamefont [1]{#1}%
\providecommand \href@noop [0]{\@secondoftwo}%
\providecommand \href [0]{\begingroup \@sanitize@url \@href}%
\providecommand \@href[1]{\@@startlink{#1}\@@href}%
\providecommand \@@href[1]{\endgroup#1\@@endlink}%
\providecommand \@sanitize@url [0]{\catcode `\\12\catcode `\$12\catcode
  `\&12\catcode `\#12\catcode `\^12\catcode `\_12\catcode `\%12\relax}%
\providecommand \@@startlink[1]{}%
\providecommand \@@endlink[0]{}%
\providecommand \url  [0]{\begingroup\@sanitize@url \@url }%
\providecommand \@url [1]{\endgroup\@href {#1}{\urlprefix }}%
\providecommand \urlprefix  [0]{URL }%
\providecommand \Eprint [0]{\href }%
\providecommand \doibase [0]{http://dx.doi.org/}%
\providecommand \selectlanguage [0]{\@gobble}%
\providecommand \bibinfo  [0]{\@secondoftwo}%
\providecommand \bibfield  [0]{\@secondoftwo}%
\providecommand \translation [1]{[#1]}%
\providecommand \BibitemOpen [0]{}%
\providecommand \bibitemStop [0]{}%
\providecommand \bibitemNoStop [0]{.\EOS\space}%
\providecommand \EOS [0]{\spacefactor3000\relax}%
\providecommand \BibitemShut  [1]{\csname bibitem#1\endcsname}%
\let\auto@bib@innerbib\@empty
\bibitem [{\citenamefont {Hasan}\ and\ \citenamefont {Kane}(2010)}]{Hasan2010}%
  \BibitemOpen
  \bibfield  {author} {\bibinfo {author} {\bibfnamefont {M. ~Z. }~\bibnamefont
  {Hasan}}\ and\ \bibinfo {author} {\bibfnamefont {C.~L.}~\bibnamefont {Kane}},\
  }\href {http://dx.doi.org/10.1103/revmodphys.82.3045} {\bibfield  {journal}
  {\bibinfo  {journal} {Rev. Mod. Phys.}\ }\textbf {\bibinfo {volume}
  {82}},\ \bibinfo {pages} {3045} (\bibinfo {year} {2010})}\ \bibinfo {note}
  {}\BibitemShut {NoStop}%
\bibitem [{\citenamefont {Qi}\ and\ \citenamefont {Zhang}(2011)}]{Qi2011}%
  \BibitemOpen
  \bibfield  {author} {\bibinfo {author} {\bibfnamefont {X.-L.}\ \bibnamefont
  {Qi}}\ and\ \bibinfo {author} {\bibfnamefont {S.-C.}\ \bibnamefont {Zhang}},\
  }\href {http://link.aps.org/doi/10.1103/RevModPhys.83.1057} {\bibfield
  {journal} {\bibinfo  {journal} {Rev. Mod. Phys.}\ }\textbf {\bibinfo {volume}
  {83}},\ \bibinfo {pages} {1057} (\bibinfo {year} {2011})}\BibitemShut
  {NoStop}%
\bibitem [{\citenamefont {Liu}\ \emph {et~al.}(2012)\citenamefont {Liu},
  \citenamefont {Zhang}, \citenamefont {Chang}, \citenamefont {Zhang},
  \citenamefont {Feng}, \citenamefont {Li}, \citenamefont {He}, \citenamefont
  {Wang}, \citenamefont {Chen}, \citenamefont {Dai}, \citenamefont {Fang},
  \citenamefont {Xue}, \citenamefont {Ma},\ and\ \citenamefont
  {Wang}}]{Liu2012}%
  \BibitemOpen
  \bibfield  {author} {\bibinfo {author} {\bibfnamefont {M.}~\bibnamefont
  {Liu}}, \bibinfo {author} {\bibfnamefont {J.}~\bibnamefont {Zhang}}, \bibinfo
  {author} {\bibfnamefont {C.-Z.}\ \bibnamefont {Chang}}, \bibinfo {author}
  {\bibfnamefont {Z.}~\bibnamefont {Zhang}}, \bibinfo {author} {\bibfnamefont
  {X.}~\bibnamefont {Feng}}, \bibinfo {author} {\bibfnamefont {K.}~\bibnamefont
  {Li}}, \bibinfo {author} {\bibfnamefont {K.}~\bibnamefont {He}}, \bibinfo
  {author} {\bibfnamefont {L.-L.}\ \bibnamefont {Wang}}, \bibinfo {author}
  {\bibfnamefont {X.}~\bibnamefont {Chen}}, \bibinfo {author} {\bibfnamefont
  {X.}~\bibnamefont {Dai}}, \bibinfo {author} {\bibfnamefont {Z.}~\bibnamefont
  {Fang}}, \bibinfo {author} {\bibfnamefont {Q.-K.}\ \bibnamefont {Xue}},
  \bibinfo {author} {\bibfnamefont {X.}~\bibnamefont {Ma}}, \ and\ \bibinfo
  {author} {\bibfnamefont {Y.}~\bibnamefont {Wang}},\ }\href
  {http://dx.doi.org/10.1103/physrevlett.108.036805} {\bibfield  {journal}
  {\bibinfo  {journal} {Phys. Rev. Lett.}\ }\textbf {\bibinfo {volume}
  {108}} (\bibinfo {year} {2012})}\ \bibinfo {note}
  {}\BibitemShut {NoStop}%
\bibitem [{\citenamefont {Checkelsky}\ \emph {et~al.}(2012)\citenamefont
  {Checkelsky}, \citenamefont {Ye}, \citenamefont {Onose}, \citenamefont
  {Iwasa},\ and\ \citenamefont {Tokura}}]{Checkelsky2012}%
  \BibitemOpen
  \bibfield  {author} {\bibinfo {author} {\bibfnamefont {J.}~\bibnamefont
  {Checkelsky}}, \bibinfo {author} {\bibfnamefont {J.}~\bibnamefont {Ye}},
  \bibinfo {author} {\bibfnamefont {Y.}~\bibnamefont {Onose}}, \bibinfo
  {author} {\bibfnamefont {Y.}~\bibnamefont {Iwasa}}, \ and\ \bibinfo {author}
  {\bibfnamefont {Y.}~\bibnamefont {Tokura}},\ }\href
  {http://www.nature.com/nphys/journal/vaop/ncurrent/abs/nphys2388.html}
  {\bibfield  {journal} {\bibinfo  {journal} {Nat Phys}\ }\textbf {\bibinfo
  {volume} {8}},\ \bibinfo {pages} {729} (\bibinfo {year} {2012})}\ \bibinfo
  {note} {}\BibitemShut {NoStop}%
\bibitem [{\citenamefont {Xu}\ \emph {et~al.}(2012)\citenamefont {Xu},
  \citenamefont {Neupane}, \citenamefont {Liu}, \citenamefont {Zhang},
  \citenamefont {Richardella}, \citenamefont {Andrew~Wray}, \citenamefont
  {Alidoust}, \citenamefont {Leandersson}, \citenamefont {Balasubramanian},
  \citenamefont {Sanchez-Barriga}, \citenamefont {Rader}, \citenamefont
  {Landolt}, \citenamefont {Slomski}, \citenamefont {Hugo~Dil}, \citenamefont
  {Osterwalder}, \citenamefont {Chang}, \citenamefont {Jeng}, \citenamefont
  {Lin}, \citenamefont {Bansil}, \citenamefont {Samarth},\ and\ \citenamefont
  {Zahid~Hasan}}]{Xu2012a}%
  \BibitemOpen
  \bibfield  {author} {\bibinfo {author} {\bibfnamefont {S.-Y.}\ \bibnamefont
  {Xu}}, \bibinfo {author} {\bibfnamefont {M.}~\bibnamefont {Neupane}},
  \bibinfo {author} {\bibfnamefont {C.}~\bibnamefont {Liu}}, \bibinfo {author}
  {\bibfnamefont {D.~M.}~\bibnamefont {Zhang}}, \bibinfo {author} {\bibfnamefont
  {A.}~\bibnamefont {Richardella}}, \bibinfo {author} {\bibfnamefont
  {L.}~\bibnamefont {Andrew~Wray}}, \bibinfo {author} {\bibfnamefont
  {N.}~\bibnamefont {Alidoust}}, \bibinfo {author} {\bibfnamefont
  {M.}~\bibnamefont {Leandersson}}, \bibinfo {author} {\bibfnamefont
  {T.}~\bibnamefont {Balasubramanian}}, \bibinfo {author} {\bibfnamefont
  {J.}~\bibnamefont {Sanchez-Barriga}}, \bibinfo {author} {\bibfnamefont
  {O.}~\bibnamefont {Rader}}, \bibinfo {author} {\bibfnamefont
  {G.}~\bibnamefont {Landolt}}, \bibinfo {author} {\bibfnamefont
  {B.}~\bibnamefont {Slomski}}, \bibinfo {author} {\bibfnamefont
  {J.}~\bibnamefont {Hugo~Dil}}, \bibinfo {author} {\bibfnamefont
  {J.}~\bibnamefont {Osterwalder}}, \bibinfo {author} {\bibfnamefont {T.-R.}\
  \bibnamefont {Chang}}, \bibinfo {author} {\bibfnamefont {H.-T.}\ \bibnamefont
  {Jeng}}, \bibinfo {author} {\bibfnamefont {H.}~\bibnamefont {Lin}}, \bibinfo
  {author} {\bibfnamefont {A.}~\bibnamefont {Bansil}}, \bibinfo {author}
  {\bibfnamefont {N.}~\bibnamefont {Samarth}}, \ and\ \bibinfo {author}
  {\bibfnamefont {M.}~\bibnamefont {Zahid~Hasan}},\ }\href
  {http://www.nature.com/nphys/journal/v8/n8/full/nphys2351.html} {\bibfield
  {journal} {\bibinfo  {journal} {Nat Phys}\ }\textbf {\bibinfo {volume} {8}},\
  \bibinfo {pages} {616} (\bibinfo {year} {2012})}\ \bibinfo {note}
  {}\BibitemShut {NoStop}%
\bibitem [{\citenamefont {Zhang}\ \emph {et~al.}(2012)\citenamefont {Zhang},
  \citenamefont {Richardella}, \citenamefont {Rench}, \citenamefont {Xu},
  \citenamefont {Kandala}, \citenamefont {Flanagan}, \citenamefont
  {Beidenkopf}, \citenamefont {Yeats}, \citenamefont {Buckley}, \citenamefont
  {Klimov}, \citenamefont {Awschalom}, \citenamefont {Yazdani}, \citenamefont
  {Schiffer}, \citenamefont {Hasan},\ and\ \citenamefont
  {Samarth}}]{Zhang2012}%
  \BibitemOpen
  \bibfield  {author} {\bibinfo {author} {\bibfnamefont {D.~M.}~\bibnamefont
  {Zhang}}, \bibinfo {author} {\bibfnamefont {A.}~\bibnamefont {Richardella}},
  \bibinfo {author} {\bibfnamefont {D.~W.}\ \bibnamefont {Rench}}, \bibinfo
  {author} {\bibfnamefont {S.-Y.}\ \bibnamefont {Xu}}, \bibinfo {author}
  {\bibfnamefont {A.}~\bibnamefont {Kandala}}, \bibinfo {author} {\bibfnamefont
  {T.~C.}\ \bibnamefont {Flanagan}}, \bibinfo {author} {\bibfnamefont
  {H.}~\bibnamefont {Beidenkopf}}, \bibinfo {author} {\bibfnamefont {A.~L.}\
  \bibnamefont {Yeats}}, \bibinfo {author} {\bibfnamefont {B.~B.}\ \bibnamefont
  {Buckley}}, \bibinfo {author} {\bibfnamefont {P.~V.}\ \bibnamefont {Klimov}},
  \bibinfo {author} {\bibfnamefont {D.~D.}\ \bibnamefont {Awschalom}}, \bibinfo
  {author} {\bibfnamefont {A.}~\bibnamefont {Yazdani}}, \bibinfo {author}
  {\bibfnamefont {P.}~\bibnamefont {Schiffer}}, \bibinfo {author}
  {\bibfnamefont {M.~Z.}\ \bibnamefont {Hasan}}, \ and\ \bibinfo {author}
  {\bibfnamefont {N.}~\bibnamefont {Samarth}},\ }\href {\doibase
  10.1103/PhysRevB.86.205127} {\bibfield  {journal} {\bibinfo  {journal} {Phys.
  Rev. B}\ }\textbf {\bibinfo {volume} {86}},\ \bibinfo {pages} {205127}
  (\bibinfo {year} {2012})}\BibitemShut {NoStop}%
\bibitem [{\citenamefont {Chang}\ \emph {et~al.}(2013)\citenamefont {Chang},
  \citenamefont {Zhang}, \citenamefont {Feng}, \citenamefont {Shen},
  \citenamefont {Zhang}, \citenamefont {Guo}, \citenamefont {Li}, \citenamefont
  {Ou}, \citenamefont {Wei}, \citenamefont {Wang}, \citenamefont {Ji},
  \citenamefont {Feng}, \citenamefont {Ji}, \citenamefont {Chen}, \citenamefont
  {Jia}, \citenamefont {Dai}, \citenamefont {Fang}, \citenamefont {Zhang},
  \citenamefont {He}, \citenamefont {Wang}, \citenamefont {Lu}, \citenamefont
  {Ma},\ and\ \citenamefont {Xue}}]{Chang2013}%
  \BibitemOpen
  \bibfield  {author} {\bibinfo {author} {\bibfnamefont {C.-Z.}\ \bibnamefont
  {Chang}}, \bibinfo {author} {\bibfnamefont {J.}~\bibnamefont {Zhang}},
  \bibinfo {author} {\bibfnamefont {X.}~\bibnamefont {Feng}}, \bibinfo {author}
  {\bibfnamefont {J.}~\bibnamefont {Shen}}, \bibinfo {author} {\bibfnamefont
  {Z.}~\bibnamefont {Zhang}}, \bibinfo {author} {\bibfnamefont
  {M.}~\bibnamefont {Guo}}, \bibinfo {author} {\bibfnamefont {K.}~\bibnamefont
  {Li}}, \bibinfo {author} {\bibfnamefont {Y.}~\bibnamefont {Ou}}, \bibinfo
  {author} {\bibfnamefont {P.}~\bibnamefont {Wei}}, \bibinfo {author}
  {\bibfnamefont {L.-L.}\ \bibnamefont {Wang}}, \bibinfo {author}
  {\bibfnamefont {Z.-Q.}\ \bibnamefont {Ji}}, \bibinfo {author} {\bibfnamefont
  {Y.}~\bibnamefont {Feng}}, \bibinfo {author} {\bibfnamefont {S.}~\bibnamefont
  {Ji}}, \bibinfo {author} {\bibfnamefont {X.}~\bibnamefont {Chen}}, \bibinfo
  {author} {\bibfnamefont {J.}~\bibnamefont {Jia}}, \bibinfo {author}
  {\bibfnamefont {X.}~\bibnamefont {Dai}}, \bibinfo {author} {\bibfnamefont
  {Z.}~\bibnamefont {Fang}}, \bibinfo {author} {\bibfnamefont {S.-C.}\
  \bibnamefont {Zhang}}, \bibinfo {author} {\bibfnamefont {K.}~\bibnamefont
  {He}}, \bibinfo {author} {\bibfnamefont {Y.}~\bibnamefont {Wang}}, \bibinfo
  {author} {\bibfnamefont {L.}~\bibnamefont {Lu}}, \bibinfo {author}
  {\bibfnamefont {X.-C.}\ \bibnamefont {Ma}}, \ and\ \bibinfo {author}
  {\bibfnamefont {Q.-K.}\ \bibnamefont {Xue}},\ } {\bibfield  {journal}
  {\bibinfo  {journal} {Science}\ }\textbf {\bibinfo {volume} {340}},\ \bibinfo
  {pages} {167} (\bibinfo {year} {2013})}\BibitemShut {NoStop}%
\bibitem [{\citenamefont {Mondal}\ \emph {et~al.}(2010)\citenamefont {Mondal},
  \citenamefont {Sen}, \citenamefont {Sengupta},\ and\ \citenamefont
  {Shankar}}]{Mondal2010}%
  \BibitemOpen
  \bibfield  {author} {\bibinfo {author} {\bibfnamefont {S.}~\bibnamefont
  {Mondal}}, \bibinfo {author} {\bibfnamefont {D.}~\bibnamefont {Sen}},
  \bibinfo {author} {\bibfnamefont {K.}~\bibnamefont {Sengupta}}, \ and\
  \bibinfo {author} {\bibfnamefont {R.}~\bibnamefont {Shankar}},\ }\href
  {http://dx.doi.org/10.1103/physrevlett.104.046403} {\bibfield  {journal}
  {\bibinfo  {journal} {Phys. Rev. Lett.}\ }\textbf {\bibinfo {volume}
  {104}},\ \bibinfo {pages} {046403} (\bibinfo {year} {2010})}\ \bibinfo
  {note} {}\BibitemShut {NoStop}%
\bibitem [{\citenamefont {Garate}\ and\ \citenamefont
  {Franz}(2010)}]{Garate2010}%
  \BibitemOpen
  \bibfield  {author} {\bibinfo {author} {\bibfnamefont {I.}~\bibnamefont
  {Garate}}\ and\ \bibinfo {author} {\bibfnamefont {M.}~\bibnamefont {Franz}},\
  }\href {http://dx.doi.org/10.1103/physrevlett.104.146802} {\bibfield
  {journal} {\bibinfo  {journal} {Phys. Rev. Lett.}\ }\textbf {\bibinfo
  {volume} {104}},\ \bibinfo {pages} {146802} (\bibinfo {year} {2010})}\
  \bibinfo {note} {}\BibitemShut {NoStop}%
\bibitem [{\citenamefont {Kong}\ \emph {et~al.}(2011)\citenamefont {Kong},
  \citenamefont {Semenov}, \citenamefont {Krowne},\ and\ \citenamefont
  {Kim}}]{Kong2011}%
  \BibitemOpen
  \bibfield  {author} {\bibinfo {author} {\bibfnamefont {B.}~\bibnamefont
  {Kong}}, \bibinfo {author} {\bibfnamefont {Y.}~\bibnamefont {Semenov}},
  \bibinfo {author} {\bibfnamefont {C.}~\bibnamefont {Krowne}}, \ and\ \bibinfo
  {author} {\bibfnamefont {K.}~\bibnamefont {Kim}},\ }\href
  {http://apl.aip.org/resource/1/applab/v98/i24/p243112_s1} {\bibfield
  {journal} {\bibinfo  {journal} {Appl. Phys. Lett.}\ }\textbf {\bibinfo
  {volume} {98}},\ \bibinfo {pages} {243112} (\bibinfo {year} {2011})}\
  \bibinfo {note} {}\BibitemShut {NoStop}%
\bibitem [{\citenamefont {Tserkovnyak}\ and\ \citenamefont
  {Loss}(2012)}]{Tserkovnyak2012}%
  \BibitemOpen
  \bibfield  {author} {\bibinfo {author} {\bibfnamefont {Y.}~\bibnamefont
  {Tserkovnyak}}\ and\ \bibinfo {author} {\bibfnamefont {D.}~\bibnamefont
  {Loss}},\ }\href {http://dx.doi.org/10.1103/physrevlett.108.187201}
  {\bibfield  {journal} {\bibinfo  {journal} {Phys. Rev. Lett.}\
  }\textbf {\bibinfo {volume} {108}},\ \bibinfo {pages} {187201} (\bibinfo
  {year} {2012})}\ \bibinfo {note}
  {}\BibitemShut {NoStop}%
\bibitem [{\citenamefont {Oroszl\'any}\ and\ \citenamefont
  {Cortijo}(2012)}]{Oroszlany2012}%
  \BibitemOpen
  \bibfield  {author} {\bibinfo {author} {\bibfnamefont {L.}~\bibnamefont
  {Oroszl\'any}}\ and\ \bibinfo {author} {\bibfnamefont {A.}~\bibnamefont
  {Cortijo}},\ }\href {\doibase 10.1103/PhysRevB.86.195427} {\bibfield
  {journal} {\bibinfo  {journal} {Phys. Rev. B}\ }\textbf {\bibinfo {volume}
  {86}},\ \bibinfo {pages} {195427} (\bibinfo {year} {2012})}\BibitemShut
  {NoStop}%
\bibitem [{\citenamefont {Wei}\ \emph {et~al.}(2013)\citenamefont {Wei},
  \citenamefont {Katmis}, \citenamefont {Assaf}, \citenamefont {Steinberg},
  \citenamefont {Jarillo-Herrero}, \citenamefont {Heiman},\ and\ \citenamefont
  {Moodera}}]{Wei2013}%
  \BibitemOpen
  \bibfield  {author} {\bibinfo {author} {\bibfnamefont {P.}~\bibnamefont
  {Wei}}, \bibinfo {author} {\bibfnamefont {F.}~\bibnamefont {Katmis}},
  \bibinfo {author} {\bibfnamefont {B.~A.}\ \bibnamefont {Assaf}}, \bibinfo
  {author} {\bibfnamefont {H.}~\bibnamefont {Steinberg}}, \bibinfo {author}
  {\bibfnamefont {P.}~\bibnamefont {Jarillo-Herrero}}, \bibinfo {author}
  {\bibfnamefont {D.}~\bibnamefont {Heiman}}, \ and\ \bibinfo {author}
  {\bibfnamefont {J.~S.}\ \bibnamefont {Moodera}},\ }\href {\doibase
  10.1103/PhysRevLett.110.186807} {\bibfield  {journal} {\bibinfo  {journal}
  {Phys. Rev. Lett.}\ }\textbf {\bibinfo {volume} {110}},\ \bibinfo {pages}
  {186807} (\bibinfo {year} {2013})}\BibitemShut {NoStop}%
\bibitem [{\citenamefont {Yang}\ \emph {et~al.}(2013)\citenamefont {Yang},
  \citenamefont {Dolev}, \citenamefont {Zhang}, \citenamefont {Zhao},
  \citenamefont {Fried}, \citenamefont {Schemm}, \citenamefont {Liu},
  \citenamefont {Palevski}, \citenamefont {Marshall}, \citenamefont {Risbud},\
  and\ \citenamefont {Kapitulnik}}]{Yang2013}%
  \BibitemOpen
  \bibfield  {author} {\bibinfo {author} {\bibfnamefont {Q.~I.}\ \bibnamefont
  {Yang}}, \bibinfo {author} {\bibfnamefont {M.}~\bibnamefont {Dolev}},
  \bibinfo {author} {\bibfnamefont {L.}~\bibnamefont {Zhang}}, \bibinfo
  {author} {\bibfnamefont {J.}~\bibnamefont {Zhao}}, \bibinfo {author}
  {\bibfnamefont {A.~D.}\ \bibnamefont {Fried}}, \bibinfo {author}
  {\bibfnamefont {E.}~\bibnamefont {Schemm}}, \bibinfo {author} {\bibfnamefont
  {M.}~\bibnamefont {Liu}}, \bibinfo {author} {\bibfnamefont {A.}~\bibnamefont
  {Palevski}}, \bibinfo {author} {\bibfnamefont {A.~F.}\ \bibnamefont
  {Marshall}}, \bibinfo {author} {\bibfnamefont {S.~H.}\ \bibnamefont
  {Risbud}}, \ and\ \bibinfo {author} {\bibfnamefont {A.}~\bibnamefont
  {Kapitulnik}},\ }\href {\doibase 10.1103/PhysRevB.88.081407} {\bibfield
  {journal} {\bibinfo  {journal} {Phys. Rev. B}\ }\textbf {\bibinfo {volume}
  {88}},\ \bibinfo {pages} {081407} (\bibinfo {year} {2013})}\BibitemShut
  {NoStop}%
\bibitem [{\citenamefont {Xiao}\ and\ \citenamefont {Chien}(1996)}]{Xiao1996}%
  \BibitemOpen
  \bibfield  {author} {\bibinfo {author} {\bibfnamefont {J.~Q.}~\bibnamefont
  {Xiao}}\ and\ \bibinfo {author} {\bibfnamefont {C.~L.}~\bibnamefont {Chien}},\
  }\href {http://dx.doi.org/10.1103/physrevlett.76.1727} {\bibfield  {journal}
  {\bibinfo  {journal} {Phys. Rev. Lett.}\ }\textbf {\bibinfo {volume}
  {76}},\ \bibinfo {pages} {1727} (\bibinfo {year} {1996})}\ \bibinfo {note}
  {}\BibitemShut {NoStop}%
\bibitem [{\citenamefont {Senapati}\ \emph {et~al.}(2011)\citenamefont
  {Senapati}, \citenamefont {Fix}, \citenamefont {Vickers}, \citenamefont
  {Blamire},\ and\ \citenamefont {Barber}}]{Senapati2011b}%
  \BibitemOpen
  \bibfield  {author} {\bibinfo {author} {\bibfnamefont {K.}~\bibnamefont
  {Senapati}}, \bibinfo {author} {\bibfnamefont {T.}~\bibnamefont {Fix}},
  \bibinfo {author} {\bibfnamefont {M.}~\bibnamefont {Vickers}}, \bibinfo
  {author} {\bibfnamefont {M.}~\bibnamefont {Blamire}}, \ and\ \bibinfo
  {author} {\bibfnamefont {Z.}~\bibnamefont {Barber}},\ }\href
  {http://dx.doi.org/10.1103/physrevb.83.014403} {\bibfield  {journal}
  {\bibinfo  {journal} {Phys. Rev. B}\ }\textbf {\bibinfo {volume} {83}},\
  \bibinfo {pages} {014403} (\bibinfo {year} {2011})}\ \bibinfo {note}
  {}\BibitemShut {NoStop}%
\bibitem [{\citenamefont {Plank}\ \emph {et~al.}(2011)\citenamefont {Plank},
  \citenamefont {Natali}, \citenamefont {Galipaud}, \citenamefont {Richter},
  \citenamefont {Simpson}, \citenamefont {Trodahl},\ and\ \citenamefont
  {Ruck}}]{Plank2011}%
  \BibitemOpen
  \bibfield  {author} {\bibinfo {author} {\bibfnamefont {N.}~\bibnamefont
  {Plank}}, \bibinfo {author} {\bibfnamefont {F.}~\bibnamefont {Natali}},
  \bibinfo {author} {\bibfnamefont {J.}~\bibnamefont {Galipaud}}, \bibinfo
  {author} {\bibfnamefont {J.}~\bibnamefont {Richter}}, \bibinfo {author}
  {\bibfnamefont {M.}~\bibnamefont {Simpson}}, \bibinfo {author} {\bibfnamefont
  {H.}~\bibnamefont {Trodahl}}, \ and\ \bibinfo {author} {\bibfnamefont
  {B.}~\bibnamefont {Ruck}},\ }\href
  {http://apl.aip.org/resource/1/applab/v98/i11/p112503_s1} {\bibfield
  {journal} {\bibinfo  {journal} {Appl. Phys. Lett.}\ }\textbf {\bibinfo
  {volume} {98}},\ \bibinfo {pages} {112503} (\bibinfo {year} {2011})}\
  \bibinfo {note} {}\BibitemShut {NoStop}%
\bibitem [{\citenamefont {Granville}\ \emph {et~al.}(2006)\citenamefont
  {Granville}, \citenamefont {Ruck}, \citenamefont {Budde}, \citenamefont
  {Koo}, \citenamefont {Pringle}, \citenamefont {Kuchler}, \citenamefont
  {Preston}, \citenamefont {Housden}, \citenamefont {Lund}, \citenamefont
  {Bittar}, \citenamefont {Williams},\ and\ \citenamefont
  {Trodahl}}]{Granville2006}%
  \BibitemOpen
  \bibfield  {author} {\bibinfo {author} {\bibfnamefont {S.}~\bibnamefont
  {Granville}}, \bibinfo {author} {\bibfnamefont {B.}~\bibnamefont {Ruck}},
  \bibinfo {author} {\bibfnamefont {F.}~\bibnamefont {Budde}}, \bibinfo
  {author} {\bibfnamefont {A.}~\bibnamefont {Koo}}, \bibinfo {author}
  {\bibfnamefont {D.}~\bibnamefont {Pringle}}, \bibinfo {author} {\bibfnamefont
  {F.}~\bibnamefont {Kuchler}}, \bibinfo {author} {\bibfnamefont
  {A.}~\bibnamefont {Preston}}, \bibinfo {author} {\bibfnamefont
  {D.}~\bibnamefont {Housden}}, \bibinfo {author} {\bibfnamefont
  {N.}~\bibnamefont {Lund}}, \bibinfo {author} {\bibfnamefont {A.}~\bibnamefont
  {Bittar}}, \bibinfo {author} {\bibfnamefont {G.}~\bibnamefont {Williams}}, \
  and\ \bibinfo {author} {\bibfnamefont {H.}~\bibnamefont {Trodahl}},\ }\href
  {http://dx.doi.org/10.1103/physrevb.73.235335} {\bibfield  {journal}
  {\bibinfo  {journal} {Phys. Rev. B}\ }\textbf {\bibinfo {volume} {73}},\
  \bibinfo {pages} {235335} (\bibinfo {year} {2006})}\ \bibinfo {note}
  {}\BibitemShut {NoStop}%
\bibitem [{\citenamefont {Richardella}\ \emph {et~al.}(2010)\citenamefont
  {Richardella}, \citenamefont {Zhang}, \citenamefont {Lee}, \citenamefont
  {Koser}, \citenamefont {Rench}, \citenamefont {Yeats}, \citenamefont
  {Buckley}, \citenamefont {Awschalom},\ and\ \citenamefont
  {Samarth}}]{Richardella2010}%
  \BibitemOpen
  \bibfield  {author} {\bibinfo {author} {\bibfnamefont {A.}~\bibnamefont
  {Richardella}}, \bibinfo {author} {\bibfnamefont {D.~M.}~\bibnamefont {Zhang}},
  \bibinfo {author} {\bibfnamefont {J.~S.}~\bibnamefont {Lee}}, \bibinfo {author}
  {\bibfnamefont {A.}~\bibnamefont {Koser}}, \bibinfo {author} {\bibfnamefont
  {D.~W.}~\bibnamefont {Rench}}, \bibinfo {author} {\bibfnamefont
  {A.~L.}~\bibnamefont {Yeats}}, \bibinfo {author} {\bibfnamefont
  {B.~B.}~\bibnamefont {Buckley}}, \bibinfo {author} {\bibfnamefont
  {D.~D.}~\bibnamefont {Awschalom}}, \ and\ \bibinfo {author} {\bibfnamefont
  {N.}~\bibnamefont {Samarth}},\ }\href {http://dx.doi.org/10.1063/1.3532845}
  {\bibfield  {journal} {\bibinfo  {journal} {Appl. Phys. Lett.}\
  }\textbf {\bibinfo {volume} {97}},\ \bibinfo {pages} {262104} (\bibinfo
  {year} {2010})}\ \bibinfo {note} {}\BibitemShut {NoStop}%
\bibitem [{\citenamefont {Wang}\ \emph {et~al.}(2011)\citenamefont {Wang},
  \citenamefont {DaSilva}, \citenamefont {Chang}, \citenamefont {He},
  \citenamefont {Jain}, \citenamefont {Samarth}, \citenamefont {Ma},
  \citenamefont {Xue},\ and\ \citenamefont {Chan}}]{Wang2011}%
  \BibitemOpen
  \bibfield  {author} {\bibinfo {author} {\bibfnamefont {J.}~\bibnamefont
  {Wang}}, \bibinfo {author} {\bibfnamefont {A.~M.}\ \bibnamefont {DaSilva}},
  \bibinfo {author} {\bibfnamefont {C.-Z.}\ \bibnamefont {Chang}}, \bibinfo
  {author} {\bibfnamefont {K.}~\bibnamefont {He}}, \bibinfo {author}
  {\bibfnamefont {J.~K.}\ \bibnamefont {Jain}}, \bibinfo {author}
  {\bibfnamefont {N.}~\bibnamefont {Samarth}}, \bibinfo {author} {\bibfnamefont
  {X.-C.}\ \bibnamefont {Ma}}, \bibinfo {author} {\bibfnamefont {Q.-K.}\
  \bibnamefont {Xue}}, \ and\ \bibinfo {author} {\bibfnamefont {M.~H.~W.}\
  \bibnamefont {Chan}},\ }\href {\doibase 10.1103/PhysRevB.83.245438}
  {\bibfield  {journal} {\bibinfo  {journal} {Phys. Rev. B}\ }\textbf {\bibinfo
  {volume} {83}},\ \bibinfo {pages} {245438} (\bibinfo {year}
  {2011})}\BibitemShut {NoStop}%
\bibitem [{\citenamefont {Chen}\ \emph {et~al.}(2013)\citenamefont {Chen},
  \citenamefont {Casu}, \citenamefont {Gajek},\ and\ \citenamefont
  {Raoux}}]{Chen_Arxiv}%
  \BibitemOpen
  \bibfield  {author} {\bibinfo {author} {\bibfnamefont {C.-T.}\ \bibnamefont
  {Chen}}, \bibinfo {author} {\bibfnamefont {E.~A.}\ \bibnamefont {Casu}},
  \bibinfo {author} {\bibfnamefont {M.}~\bibnamefont {Gajek}}, \ and\ \bibinfo
  {author} {\bibfnamefont {S.}~\bibnamefont {Raoux}},\ }\href@noop {}
  {\bibfield  {journal} {\bibinfo  {journal} {ArXiv e-prints}\ } (\bibinfo
  {year} {2013})},\ \Eprint {http://arxiv.org/abs/1303.3325} {arXiv:1303.3325
  [cond-mat.mes-hall]} \BibitemShut {NoStop}%
\bibitem [{\citenamefont {Lu}\ \emph {et~al.}(2011)\citenamefont {Lu},
  \citenamefont {Shi},\ and\ \citenamefont {Shen}}]{Lu2011a}%
  \BibitemOpen
  \bibfield  {author} {\bibinfo {author} {\bibfnamefont {H.-Z.}\ \bibnamefont
  {Lu}}, \bibinfo {author} {\bibfnamefont {J.}~\bibnamefont {Shi}}, \ and\
  \bibinfo {author} {\bibfnamefont {S.-Q.}\ \bibnamefont {Shen}},\ }\href
  {http://dx.doi.org/10.1103/physrevlett.107.076801} {\bibfield  {journal}
  {\bibinfo  {journal} {Phys. Rev. Lett.}\ }\textbf {\bibinfo {volume}
  {107}},\ \bibinfo {pages} {076801} (\bibinfo {year} {2011})}\ \bibinfo
  {note} {}\BibitemShut {NoStop}%
  \bibitem[{\citenamefont{Hikami et~al.}(1980)\citenamefont{Hikami, Larkin, and
  Nagaoka}}]{Hikami1980}
\bibinfo{author}{\bibfnamefont{S.}~\bibnamefont{Hikami}},
  \bibinfo{author}{\bibfnamefont{A.~I.} \bibnamefont{Larkin}},
  \bibnamefont{and} \bibinfo{author}{\bibfnamefont{Y.}~\bibnamefont{Nagaoka}},
  \bibinfo{journal}{Prog. Theor. Phys.} \textbf{\bibinfo{volume}{63}},
  \bibinfo{pages}{707} (\bibinfo{year}{1980}).
\end{thebibliography}

%

\end{document}